\documentclass[12pt]{article}

\usepackage{amsmath,amssymb,setspace,enumitem,soul,color,url,float}
\usepackage{tikz-cd}

\begin{document}

\begin{center}
\textbf{\Large Generalised regression estimation given\\ imperfectly matched auxiliary data} \\
\emph{Li-Chun Zhang}\footnote{S3RI/University of Southampton (email: L.Zhang@soton.ac.uk), Statistisk sentralbyr\aa\ \& University of Oslo.}
\end{center}

\begin{abstract} Generalised regression estimation allows one to make use of available auxiliary information in survey sampling. We develop three types of generalised regression estimator when the auxiliary data cannot be matched perfectly to the sample units, so that the standard estimator is inapplicable. The inference remains design-based. Consistency of the proposed estimators is either given by construction or else can be tested given the observed sample and links. Mean square errors can be estimated. A simulation study is used to explore the potentials of the proposed estimators. 
\end{abstract}

\noindent
\emph{Keywords:} record linkage, incidence weights, reverse incidence weights

\section{Introduction}

Let $\{ y_i : i\in s\}$ be the values observed in a sample $s$ from the population $U = \{ 1, ..., N\}$.  Design-unbiased estimation of the population total $Y = \sum_{i=1}^N y_i$ can be achieved using the sample inclusion probabilities $\pi_i = \mbox{Pr}(i\in s)$ for $i\in U$. Let $x_i$ be the vector of known auxiliary values for each $i\in U$. By incorporating these auxiliary values, the \emph{generalised regression (GREG)} estimator (e.g. S\"{a}rndal, 1992) can often improve the efficiency of estimation. The GREG estimator of $Y$ is given by
\begin{equation} \label{GREG}
\widehat{Y}_{GR} = \widehat{Y} + (X - \widehat{X})^{\top} b = X^{\top} b + \sum_{i\in s} (y_i - x_i^{\top} b)/\pi_i 
\end{equation}
where $\widehat{Y} = \sum_{i\in s} y_i/\pi_i$ is the Horvitz-Thompson (HT) estimator of $Y$ and $\widehat{X}$ that of $X = \sum_{i\in U} x_i$, and $b = \big( \sum_{i\in s} c_i x_i x_i^{\top}/\pi_i \big)^{-1} \big( \sum_{i\in s} c_i x_i y_i/\pi_i  \big)$ is
a weighted least-squares estimate of the coefficients of a linear regression model of $y_i$ on $x_i$. The constants $c_i$ can be introduced given heterogeneous regression errors; it is also common to set $c_i \equiv 1$. 

To calculate the GREG estimator \eqref{GREG}, one needs the $(x,y)$-values for each sample unit. However, it may be impossible to match the sample and the auxiliary database perfectly, because one does not have a common, unique identifier in both sources. Record linkage based on linkage \emph{key} variables (e.g. Fellegi and Sunter, 1969; Herzog et al., 2007; Christen, 2012; Harron et al., 2015), such as name and birth date, will be imperfect if some of them are incorrectly recorded in either source, so that any pairing of $(x,y)$ by a link may not actually refer to the same unit. This causes the problem for GREG estimation in situations where the auxiliary data cannot be perfectly matched.

While exists a growing literature on linear regression analysis based on linked datasets --- see e.g. Lahiri and Larsen (2005),  Chambers (2009), Chambers and Da Silva (2020) and Zhang and Tuoto (2020) under the frequentist framework of inference, our perspective is different here. The interest is not the regression relationship itself. The aim is to utilise the auxiliary information to improve the efficiency of population total (or mean) estimation, where the linear model plays the role of an assisting model (e.g. S\"{a}rndal, 1992), and the inference is based on the sampling design rather than the model. For instance, in regression analysis the auxiliary population total is of little consequence, whereas it is of central importance to GREG estimation, and the ostensible total of the $x$-values in the auxiliary database, denoted by $X_A$ over $A$, cannot be used directly when the matching between $A$ and $U$ is \emph{incomplete}, i.e. they are not one-one matched in truth.

To the best of our knowledge, Breidt et al. (2018) is the only previous work that addresses the problem from our perspective. In their motivating example, the population $U$ consists of recreational fishing boat trips along the Atlantic Coast of South Carolina each year, and the $y$-value is the catch on each boat trip. The auxiliary database consists of the boat's logbook (including data of catch) required to be reported to the South Carolina Department of Natural Resources. The quality of record linkage is rather poor, and one cannot be sure that all the trips are reported in the logbooks. Breidt et al. (2018) consider a \emph{difference} estimator, which makes use of multiple links for the sample trips. Although the estimator is biased, one is able to reduce the mean squared error (MSE) of estimation. The difference predictor is a special case of GREG predictor given fixed regression coefficients. As these authors point out, there is a need for developing methods which allow the predictor to be estimated from the sample actually observed. 

We shall develop three types of GREG estimators and their approximate variance estimators, when the matching between the population and the auxiliary database is incomplete and record linkage between them is imperfect. The conditions for design-consistent estimators are specified, which can be tested given the observed sample and links, if the conditions cannot be verified directly based on linking the entire population and auxiliary database. Thus, the MSE of estimation can be estimated.  

In Section \ref{ambiguity} we outline the underlying linkage structure of the problem and the inference framework. The GREG-estimators are developed in Section \ref{estimators}. A simulation study is used to study the relative merits of the proposed estimators in Section \ref{simulation}, also in comparison to the HT estimator that ignores the auxiliary information and the hypothetical ideal GREG estimator. Some conclusions and final remarks are given in Section \ref{final}.

\section{Entity ambiguity and inference framework} \label{ambiguity}

Imperfect matching between separate data files arises from the ambiguity surrounding the set of unique entities underlying these data files. Record linkage, or entity resolution, results in one or several \emph{links} between a record in one file and the records in another. A link between a pair of records is a \emph{match} if the records refer to the same entity, the link is \emph{false} otherwise. False links and missed matches are caused by errors of the key variables used for record linkage, in the absence of a true identifier (i.e. a perfect key variable). While the formulation can be extended to include duplicated records in the same file, we shall assume that duplicates are absent in the following. 

Denote by $M = \{ (i, \iota_i)\}$ the matches between the population $U$ and the auxiliary database $A$,  where $\iota_i$ in $A$ is the matching record of $i$ in $U$. Let $N_A = |A|$ be the size of $A$, which may differ to $N = |U|$, e.g. if $U$ and $A$ are not one-one correspondent in terms of the matches. Let the population (set of) links be given as
\[
L = \bigcup_{i\in U} i\times \alpha_i = \bigcup_{\ell\in A} \beta_{\ell} \times \ell
\]
where $\alpha_i$ contains the records in $A$ linked to unit $i \in U$, with cardinality $d_i = |\alpha_i| \geq 1$, and $\beta_{\ell}$ contains the units in $U$ that are linked to record $\ell \in A$, and $\beta_{\ell}$ may be empty for some $\ell$. Let $m_{\ell} = |\beta_{\ell} |$ be the cardinality of $\beta_{\ell}$, where $m_{\ell} = 0$ if $\beta_{\ell}$ is empty.

Some explanations are needed for this set-up. In a situation where one is able to link $U$ and $A$, one can easily impose a restriction that any record $\ell \in A$ is linked to at least one unit in $U$ as well. However, in practice, it is often the case that one is only able to link the sample units in $s$ to $A$, for $s\subset U$, but not the records in $A$ to $U\setminus s$, because the key variables are only observed in $s$ but not $U\setminus s$. This is indeed the situation considered by Breidt et al. (2018). Thus, to ensure general applicability, we assume that the \emph{direction} of linkage is from $U$ to $A$, which allows one to ensure that each unit $i\in U$ is linked to at least one record in $A$, no matter how likely (or unlikely) one judges that a link may be correct. That is, for any given $i\in U$, one finds one or more records in $A$ that can be linked to it, but one does not look for the units in $U$ that can be linked to any given record $\ell \in A$. It follows that $\beta_{\ell}$ may be empty for some records in $A$. (Of course, the methods developed in this paper remain applicable if all $\beta_{\ell}$ are non-empty.)   

Given the population links $L$ from $U$ to $A$, $\alpha_i$ is fully observed for any sample unit in $s$, as well as the sample links $L_s = \bigcup_{i\in s} i\times \alpha_i \subset L$. Whereas $\beta_{\ell}$ for any $\ell \in A$ may not be fully observed in $L_s$, based on which one only observes $s_{\ell} = s\cap \beta_{\ell}$ for $\ell \in \alpha(s) = \bigcup_{i\in s} \alpha_i$. The example below provides an illustration.

\paragraph{\em Example} Let $N =6$ and $N_A = 6$. The records in $A$ are enumerated $\ell = 1, ..., 6$ as they are known in $A$, and $\iota_1, ..., \iota_5$ in parentheses according to their unknown matches in $U$, where the matches are shown as dashed lines. Notice that $U$ and $A$ are not one-one correspondent in terms of the matches, despite $N=N_A$. The population unit $i=6$ in $U$ is an \emph{unmatched} unit and the record $\ell =6$ presents an unmatched entity in $A$. The population links $L$ are given by the solid arrows. 
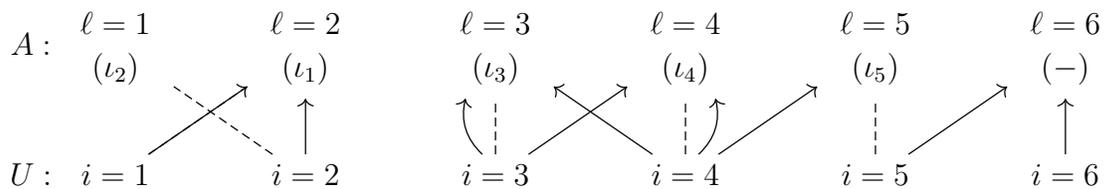
\begin{center}
\begin{tikzcd}[cramped]
A:\hspace{-13mm} & \begin{array}{c} \ell=1\\ (\iota_2) \end{array} & \begin{array}{c} \ell=2\\ (\iota_1) \end{array} & 
\begin{array}{c} \ell=3\\ (\iota_3) \end{array} & \begin{array}{c} \ell=4\\ (\iota_4) \end{array} 
& \begin{array}{c} \ell=5\\ (\iota_5) \end{array} & \begin{array}{c} \ell=6\\ (-) \end{array} \\
U:\hspace{-13mm}& i=1 \arrow[ur, dash, dashed] \arrow[ur] & i=2 \arrow[ul, dash, dashed] \arrow[u] 
& i=3 \arrow[u, dash, dashed] \arrow[u, bend left] \arrow[ur] & i=4 \arrow[u, dash, dashed] \arrow[u, bend right] \arrow[ul] \arrow[ur] 
& i=5 \arrow[u, dash, dashed] \arrow[ur] & i=6 \arrow[u]
\end{tikzcd}
\end{center}
Let the sample be $s = \{ 2, 3, 4\}$ from $U$. The sample links are 
\[
L_s = \{ (2, \iota_1), (3, \iota_3), (3, \iota_4), (4, \iota_3), (4, \iota_4), (4, \iota_5)\}
\] 
such that $\alpha_2 = \{ \iota_1\}$, $\alpha_3 = \{ \iota_3, \iota_4 \}$ and $\alpha_4 = \{ \iota_3, \iota_4, \iota_5 \}$. These are fully observed in $L_s$. Moreover, we observe $s_{\ell}$ for $\ell \in \alpha(s) = \{ \iota_1, \iota_3, \iota_4, \iota_5\}$, where $s_{\iota_1} = \{ 2\}$, $s_{\iota_3} = s_{\iota_4} = \{ 3, 4 \}$ and $s_{\iota_5} = \{ 4 \}$. However, any observed $s_{\ell}$ can differ from $\beta_{\ell}$, since it is possible for other units in $U\setminus s$ to be linked to the records in $\alpha(s)$, such as $\beta_{\iota_1} = \{ 1, 2\} \neq s_{\iota_1}$. Finally, for these sample units, the missing match is $(2, \iota_2)$ for $i=2$; the false links are $(2, \iota_1)$ for $i=2$, $(3, \iota_4)$ for $i=3$, and $(4, \iota_3)$ and $(4, \iota_5)$ for $i=4$. $\square$

\bigskip
Generally, in the presence of entity ambiguity, we have $L \neq M$, where the false links are $L\setminus M$, and the missing matches are $M\setminus L$. For the methods of GREG estimation given imperfectly matched auxiliary data and the associated uncertainty assessment, we shall treat $(U, A)$ and all the associated the linkage key variables as \emph{fixed}, whatever the underlying mechanism that has generated the key-variable errors and the chosen linkage method. Hence, the population links $L$ are fixed as well. The expectation and variance of an estimator will be evaluated only with respect to the sampling design.

\section{Estimators} \label{estimators}

We consider two settings: (I) linkage from $U$ to $A$ is possible and $L$ is known, (II) linkage is only possible from $s$ to $A$ and one observes only $L_s$ associated with $s$. Below we first consider a class of estimators that are only feasible under the first setting, and then two classes of estimators that are feasible under both settings.

\subsection{Setting-I: given $L$} 

We observe fully $\beta_{\ell}$ for any $\ell \in \alpha(s)$, since $L$ is known. Let $\omega_{i\ell}$ be the \emph{incidence weights} that are non-negative constants of sampling, where $\omega_{i\ell} = 0$ for $i\not \in \beta_{\ell}$, and
\begin{equation} \label{wp}
\sum_{i\in \beta_{\ell}} \omega_{i\ell} = 1
\end{equation} 
In the special case of $\omega_{i\ell} \equiv 1/m_{\ell}$ for $i\in \beta_{\ell}$, the weights are referred to as the \emph{multiplicity weights}. One can vary $\omega_{i\ell}$ subjected to the constraint \eqref{wp}, e.g. based on the comparison scores used for record linkage (Fellegi and Sunter, 1969). In any case, the weights are constants of sampling given $U$, $A$, $L$ and the associated key variables. 

Let $z_i$ be the constant auxiliary value for each $i\in U$, which is given by
\[
z_i = \sum_{\ell \in \alpha_i} \omega_{i\ell} x_{\ell}
\]
Notice that we have $Z = \sum_{i\in U} z_i = \sum_{\ell \in A} x_{\ell} =X_A$, if $m_{\ell} >0$ for all $\ell \in A$, since
\[
Z = \sum_{i\in U} z_i = \sum_{i\in U} \sum_{\ell\in \alpha_i} \omega_{i\ell} x_{\ell} = \sum_{\ell\in A} x_{\ell} \sum_{i\in \beta_{\ell}} \omega_{i\ell} = X_A
\] 
by virtue of \eqref{wp}. However, we do not assume this to be the case generally.
For an illustration using Example earlier, given the sample $s = \{ 2, 3, 4\}$, we have $z_2 = \omega_{22}  x_2$ for $i=2$ where $\omega_{12} +\omega_{22} = 1$ for $\ell =2$, and $z_3 = \omega_{33} x_3 + \omega_{34} x_4$ and $z_4 = \omega_{43} x_3 + \omega_{44} x_4 + \omega_{45} x_5$, where $\omega_{33} + \omega_{43} = 1$ and $\omega_{34} + \omega_{44} = 1$ and $\omega_{45} = 1$. In particular, the multiplicity weights are given by
$\omega_{22} = \omega_{33} = \omega_{43} = \omega_{34} = \omega_{44} = 1/2$, since $\beta_2$, $\beta_3$ and $\beta_4$ are all of size 2. The population total $Z$ is given by $Z = \sum_{\ell=2}^6 x_{\ell} = X_A - x_1$. 

We observe the population total $Z = \sum_{i\in U} z_i$ given $L$ and $\{ x_{\ell} : \ell\in A\}$. A \emph{population incidence (PI)} GREG estimator can be given by \eqref{GREG} based on $z_i$ instead of $x_i$, i.e.
\begin{equation} \label{PI}
\widehat{Y}_z = \widehat{Y} + (Z - \widehat{Z})^{\top} b_z = Z^{\top} b_z + \sum_{i\in s} (y_i - z_i^{\top} b_z)/\pi_i 
\end{equation}
where $b_z = \big( \sum_{i\in s} c_i z_i z_i^{\top}/\pi_i \big)^{-1} \big( \sum_{i\in s} c_i z_i y_i/\pi_i  \big)$. Let $B_z = \big( \sum_{i\in U} c_i z_i z_i^{\top} \big)^{-1} \big( \sum_{i\in U} c_i z_i y_i \big)$. The variance of the PI-GREG estimator is approximately given by that of 
\[
\hat{e}_z = \sum_{i\in s} e_{iz}/\pi_i \qquad\text{where}\quad e_{iz} = y_i - z_i^{\top} B_z 
\]
The estimator \eqref{PI} is design-consistent as $n,N\rightarrow \infty$, provided the ideal GREG estimator \eqref{GREG} is consistent. This is a main advantage that $L$ and $Z$ are known under setting-I.

\subsection{Setting-II: given $L_s$} 

Suppose only $L_s$ is observed over $s\times \alpha(s)$, where $L_s \subset L$. Since we observe only $s_{\ell}$ but not necessarily $\beta_{\ell}$ for any $\ell \in \alpha(s)$, the incidence weights by \eqref{wp} are unknown. This is the situation considered by Breidt et al. (2018), who set $\omega_{i\ell}$ heuristically according to the assessed quality of the links in $L_s$. Now, provided $M\setminus L = \emptyset$, i.e. all the matches are among the links in $L$ although one does not know them all, one may let 
\[
\omega_{i\ell} = \mbox{Pr}\big[ (i,\ell)\in M | \beta_{\ell}\neq \emptyset \big]
\]
be the probability that a link $(i,\ell)$ is the match for $\ell$, so that the constraint \eqref{wp} is satisfied. However, one still would not know the total $Z$ of the corresponding $\{ z_i : i\in U\}$, as long as $L$ is unknown. Moreover, the probability above cannot be calculated correctly for all $i\in s_{\ell}$ without knowing the other links $(\beta_{\ell}\setminus s_{\ell})\times \ell$, even if the error mechanism of the key-variables were known. Therefore, this is not a viable option. Below we consider two types of estimators, where $\omega_{i\ell}$ is fully determined given the observed $\alpha_i$ for any $i\in s$.

\subsubsection{Reverse incidence weights}

Let the \emph{reverse incidence weights} be such that, for each $i\in U$, we have
\begin{equation} \label{ws}
\sum_{\ell \in \alpha_i} \omega_{i\ell} = 1
\end{equation}
While the incidence weights \eqref{wp} sum to one for any record $\ell$ in $A$ with $m_{\ell} >0$, the weights \eqref{ws} sum to one in the opposite direction over $\alpha_i$ for any unit $i\in U$. Hence, the adjective reverse. While the incidence weights require the knowledge of the population links $L$, the reverse incidence weights are always available given the sample links $L_s$. 

Let $x_{i\omega} = \sum_{\ell \in \alpha_i} \omega_{i\ell} x_{\ell}$ be constructed $x$-value of $i\in U$ based on the reverse incidence weights. Again, one may define the weights according to the relative plausibility of the links in $i\times \alpha_i$ based on record linkage. The weights are then constants of sampling given $U$, $A$, $L$ and the associated key variables. Let $X_{\omega} = \sum_{i\in U} x_{i\omega}$. When $L$ is known, a \emph{population reverse incidence (PRI)} GREG estimator can be given as
\[
\widetilde{Y}_{\omega} = X_{\omega}^{\top} b_{\omega} + \sum_{i\in s} (y_i - x_{i\omega}^{\top} b_{\omega})/\pi_i 
= \widehat{Y} + (X_{\omega} - \widehat{X}_{\omega})^{\top} b_{\omega}
\]
where $\widehat{X}_{\omega} = \sum_{i\in s} x_{i\omega}/\pi_i$ and $b_{\omega} = \big( \sum_{i\in s} c_i x_{i\omega} x_{i\omega}^{\top}/\pi_i \big)^{-1} \big( \sum_{i\in s} c_i x_{i\omega} y_i/\pi_i  \big)$. The variance of the PRI-GREG estimator is approximately given by that of 
\[
\hat{e}_{\omega} = \sum_{i\in s} e_{i\omega}/\pi_i \quad\text{where}\quad e_{i\omega} = y_i - x_{i\omega}^{\top} B_{\omega} 
\quad\text{and}\quad 
B_{\omega} = \big( \sum_{i\in U} c_i x_{i\omega} x_{i\omega}^{\top} \big)^{-1} \big( \sum_{i\in U} c_i x_{i\omega} y_i \big)
\]

For the general case where $X_{\omega}$ is unknown because $L$ is unknown. The \emph{sample reverse incidence (SRI)} GREG estimator of $Y$ is given as
\begin{equation} \label{SRI}
\widehat{Y}_{\omega} = N\bar{X}_A^{\top} b_{\omega} + \sum_{i\in s} (y_i - x_{i\omega}^{\top} b_{\omega})/\pi_i 
= \widehat{Y} + (N \bar{X}_A - \widehat{X}_{\omega})^{\top} b_{\omega}
\end{equation}
where $\bar{X}_A = X_A/N_A$ is the mean of the $x$-values over $A$. Writing $\widehat{Y}_{\omega} = \sum_{i\in s} w_i y_i$ as a linear estimator with the sample weights $\{ w_i : i\in s\}$, we have $\sum_{i\in s} w_i x_{i\omega} = N \bar{X}_A$. The SRI-GREG estimator has the same approximate variance as the PRI-GREG estimator, since the first-order approximations of the two only differ by a constant $N(\bar{X}_A - \bar{X}_{\omega})^{\top} B_{\omega}$ where $\bar{X}_{\omega} = X_{\omega}/N$. However, insofar as $\bar{X}_{\omega} \neq \bar{X}_A$, the estimator \eqref{SRI} will be biased under repeated sampling. An additional assumption is needed for design-consistency, i.e. 
\begin{equation} \label{addition}
\lim_{N\rightarrow \infty} (\bar{X}_{\omega} - \bar{X}_A) = 0
\end{equation}
Intuitively, this assumption may seem reasonable, as long as the errors of the linkage key variables associated with $U$ and $A$ are unrelated to the $x$-values in $A$. A more detailed condition will be given later in Section \ref{LGREG}. For the moment, notice that in practice the assumption can be tested based on the observed statistic $\widehat{\bar{X}}_{\omega} - \bar{X}_A$.

A special case of the reverse incidence weights is worth mentioning. For each $i\in U$, let $\ell_i$ be the \emph{best} link among all $\ell \in \alpha_i$. Let $\omega_{i\ell_i} = 1$ and $\omega_{i\ell} =0$ for any other $\ell \in \alpha_i$. It should be pointed out that this is not a special case of incidence weights \eqref{wp}, since $\ell_i$ is chosen among $\alpha_i$ not $\beta_{\ell_i}$ and it is possible for any given $\ell \in A$ to be the best link for more than one unit in $U$. Let $x_i^* = x_{\ell_i}$ be the best-link $x$-value of $i\in s$. This can be relevant for secondary users, who are only given these best-link auxiliary values, but have no access to the other links because the linkage is performed by another party. 

Let $X^* = \sum_{i\in U} x_i^*$. The \emph{sample best-link (SBL)} GREG estimator of $Y$ is given as
\begin{equation} \label{SBL}
\widehat{Y}_b = N\bar{X}_A^{\top} b^* + \sum_{i\in s} (y_i - x_i^{*\top} b^*)/\pi_i = \widehat{Y} + (N \bar{X}_A - \widehat{X}^*)^{\top} b^*
\end{equation}
where $\widehat{X}^* = \sum_{i\in s} x_i^*/\pi_i$ and $b^* = \big( \sum_{i\in s} c_i x_i^* x_i^{*\top}/\pi_i \big)^{-1} \big( \sum_{i\in s} c_i x_i^* y_i/\pi_i  \big)$. The variance of the SBL-GREG estimator is approximately given by that of 
\[
\hat{e}^* = \sum_{i\in s} e_i^* /\pi_i \quad\text{where}\quad e_i^* = y_i - x_i^{*\top} B^* \quad\text{and}\quad
B^* = \big( \sum_{i\in U} c_i x_i^* x_i^{*\top} \big)^{-1} \big( \sum_{i\in U} c_i x_i^* y_i \big)
\]
The estimator \eqref{SBL} is design-consistent, provided $X^*/N - \bar{X}_A \rightarrow 0$, as $N\rightarrow \infty$, and the assumption can be tested based on the observed statistic $\widehat{X}^*/N - \bar{X}_A$.

\subsubsection{GREG over $L_s$} \label{LGREG}

Let $N_L = \sum_{i\in U} d_i = |L|$ be the number of links in $L$. Let $X_L = \sum_{i\in U} \sum_{\ell \in \alpha_i} x_{\ell}$ and $\bar{X}_L = X_L/N_L$ be the total and mean of $x$ over $L$, respectively. Let $B_L$ be a vector of coefficients of the same dimension as $X_L$. Let
\[
\widetilde{Y}_L = \widehat{Y} + N (\bar{X}_A - \widehat{\bar{X}}_L)^{\top} B_L
\]
where $\widehat{\bar{X}}_L = \widehat{X}_L/\widehat{N}_L$, and $\widehat{X}_L = \sum_{i\in s} \sum_{\ell\in \alpha_i} x_{\ell}/\pi_i$, and $\widehat{N}_L =  \sum_{i\in s} \sum_{\ell\in \alpha_i} 1/\pi_i$. Clearly, $\widetilde{Y}_L$ is design-consistent for $Y$ if $\lim_{N\rightarrow \infty} (\bar{X}_L - \bar{X}_A) = 0$, which can be tested based on the observed statistic $\widehat{\bar{X}}_L - \bar{X}_A$. Below  
we consider first this condition in more details, and then the estimation of $B_L$ given the sample links. $L_s$.

Let $\{ \omega_{i\ell} : (i,\ell)\in L\}$ be the incidence weights \eqref{wp} associated with $L$. Let $A_L$ be the link-projection of $U$ onto $A$, containing the $N_{AL}$ linked records in $A$. Let $a_{\ell} =1$ if $\ell\in A_L$ and 0 otherwise. We assume $N/N_L = O(1)$ and $N_{AL}/N_A = O(1)$, as $N\rightarrow \infty$. We say $\omega_{i\ell}$ and $a_{\ell}$ are \emph{non-informative} of the $x$-values asymptotically, as $N\rightarrow \infty$, if
\begin{equation} \label{NPA}
\begin{cases}
\lim\limits_{N\rightarrow \infty} \frac{\sum_{(i,\ell)\in L} \omega_{i\ell} x_{\ell}}{N_L} - \frac{\sum_{(i,\ell)\in L} \omega_{i\ell}}{N_L} \bar{X}_L = 0 \\
\lim\limits_{N\rightarrow \infty} \frac{\sum_{\ell\in A} a_{\ell} x_{\ell}}{N_A} - \frac{\sum_{\ell\in A} a_{\ell}}{N_A} \bar{X}_A
\end{cases}
\end{equation}
In other words, as $N \rightarrow \infty$, the empirical covariance of $\omega_{i\ell}$ and $x_{\ell}$ over $L$ tends to 0, as well as that of $a_{\ell}$ and $x_{\ell}$ over $A$. We have then $\lim_{N\rightarrow \infty} (X_{AL}/N_{AL} - \bar{X}_L) =0$, since $\sum_{(i,\ell)\in L} \omega_{i\ell} = N_{AL}$ and $\sum_{(i,\ell)\in L} \omega_{i\ell} x_{\ell} = X_{AL} = \sum_{\ell \in A_L} x_{\ell}$ in the first part of \eqref{NPA}, and $\lim_{N\rightarrow \infty} (X_{AL}/N_{AL} - \bar{X}_A) =0$ due to the second part of \eqref{NPA}. It follows that $\bar{X}_L - \bar{X}_A \rightarrow 0$, as $N\rightarrow \infty$, and the estimator $\widetilde{Y}_L$ above is consistent. 

Moreover, we have $\lim_{N\rightarrow \infty} (\bar{X}_{\omega} - \bar{X}_L) =0$, if the first part of \eqref{NPA} holds when $\omega_{i\ell}$ are the reverse incidence weights \eqref{ws}, where $\sum_{(i,\ell)\in L} \omega_{i\ell} = N$ and $\sum_{(i,\ell)\in L} \omega_{i\ell} x_{\ell} = X_{\omega}$. Thus, the condition \eqref{addition} for the SRI-GREG estimator \eqref{SRI} is satisfied if the reverse incidence weights are non-informative of the $x$-values in addition to \eqref{NPA}. In reality, requiring the first part of \eqref{NPA} to hold for both types of weights is essentially the same as requiring it to hold for either type of weights, since it is hard to imagine a situation where the condition holds only for one type of weights but not the other type.

To reveal the estimator of $B_L$, we observe that
\[
\widetilde{Y}_L = N \bar{X}_A^{\top} B_L + \sum_{i\in s} \frac{1}{\pi_i} 
\Big( \sum_{\ell \in \alpha_i}  \omega_{i\ell}  y_i - \frac{N}{\widehat{N}_L} \sum_{\ell \in \alpha_i} x_{\ell}^{\top} B_L \Big)
\]
given any reverse incidence weights \eqref{ws}. Thus, $B_L$ can be set according to GREG of $\omega_{i\ell} y_i$ on $x_{\ell}/\hat{r}$
over $L_s$, where $\hat{r} = \widehat{N}_L/N$. A \emph{sample link-set (SLS)} GREG estimator of $Y$ is
\begin{equation} \label{SLS}
\widehat{Y}_L = N\bar{X}_A^{\top} b_L + \sum_{i\in s} (y_i - x_{iL}^{\top} b_L/\hat{r})/\pi_i 
= \widehat{Y} + N (\bar{X}_A - \widehat{\bar{X}}_L)^{\top} b_L
\end{equation}
where $x_{iL} = \sum_{\ell \in \alpha_i} x_{\ell}$, and 
$b_L = \hat{r} \big( \sum_{i\in s} \sum_{\ell\in \alpha_i} c_{i\ell} x_{\ell} x_{\ell}^{\top}/\pi_i \big)^{-1} 
\big( \sum_{i\in s} \sum_{\ell\in \alpha_i} c_{i\ell} x_{\ell} \omega_{i\ell} y_i/\pi_i  \big)$, 
since $\pi_i$ is the inclusion probability of a link $(i,\ell)$ in $L_s$. The target of $b_L$ over sampling is  
$B_L = r \big( \sum_{i\in U} \sum_{\ell\in \alpha_i} c_{i\ell} x_{\ell} x_{\ell}^{\top} \big)^{-1} 
\big( \sum_{i\in U} \sum_{\ell\in \alpha_i} c_{i\ell} x_{\ell} \omega_{i\ell} y_i \big)$, where $r = N_L/N$. 

The first-order Taylor expansion of the SLS-GREG estimator \eqref{SLS} is given by
\[
\widehat{Y}_L \doteq (\bar{X}_A - \bar{X}_L)^{\top} b_L + 
\widehat{Y} - r^{-1} \widehat{X}_L^{\top} B_L + r^{-1} \bar{X}_L^{\top} B_L \widehat{N}_L 
= (\bar{X}_A - \bar{X}_L)^{\top} b_L + \sum_{i\in s} e_{iL}' /\pi_i 
\]
where $e_{iL}' = e_{iL} + d_i/r$, and $e_{iL} = y_i - x_{iL}^{\top} B_L/r$ is the sum of population link-set GREG residuals over $\alpha_i$, for $i\in U$. The extra term $d_i/r$ arises from the estimator $\widehat{N}_L$. Given \eqref{NPA}, an approximate variance estimator can be that of $\sum_{i\in s} e_{iL}'/\pi_i$.

\subsection{Relative efficiency}

Of the three types of GREG estimators above, the PI-GREG estimator \eqref{PI} is based on the incidence weights \eqref{wp}, the SRI-GREG estimator \eqref{SRI} and the SLS-GREG estimator \eqref{SLS} are based on the reverse incidence weights \eqref{ws}; the first two are based on GREG over $s$, and the last one is based on GREG over $L_s$. A key factor to the relative efficiency is the covariance between the dependent and independent variables of the regression. 

Consider the simple linear regression model as the assisting model, where the model covariance $(y_i, x_i)$ is a scaler that is easy to comprehend. The PI-GREG estimator depends on the covariance between $y_i$ and $z_i = \sum_{\ell \in \alpha_i} \omega_{i\ell} x_{\ell}$, the SRI-GREG estimator on that between $y_i$ and $x_{i\omega} = \sum_{\ell \in \alpha_i} \omega_{i\ell} x_{\ell}$, and the SLS-GREG on that between $\omega_{i\ell} y_i$ and $x_{\ell}$. For any given $i\in U$ and choice of $\omega_{i\ell}$ over $\alpha_i$, we have
\[
\mbox{Cov}(y_i, \omega_{i\ell} x_{\ell}) = \mbox{Cov}(\omega_{i\ell} y_i, x_{\ell}) 
= \begin{cases} \omega_{i\ell} \mbox{Cov}(y_i, x_i) & \text{if } \ell =\iota_i\\ 0 & \text{if } \ell_i\neq \iota_i \end{cases}
\]
where $Cov(y_i, x_i)$ refers to the model covariance between $y_i$ and the matched $x_i$, and $Cov(y_i, x_j) =0$ for two different units $i\neq j$. Thus, given the presence of false links and the fact that $\omega_{i\ell} \in [0,1]$, the model covariance is reduced given imperfectly matched auxiliary values, and the population GREG slope coefficient will be attuned towards 0 compared to that given matched auxiliary values. This is the main reason why GREG estimation given imperfectly matched auxiliary values will lose efficiency compare to the ideal situation where the matches are known.  

Meanwhile, what is important in practice is whether using the constructed auxiliary values based on the sample links $L_s$ can still improve the efficiency compared to the HT-estimator that ignores the auxiliary information altogether. It is possible to equate the HT-estimator with the GREG estimator that uses an intercept-only assisting model, where the covariance between $y_i$ and the constant independent variable $x_i \equiv 1$ is zero by definition. Using either the incidence weights or the reverse incidence weights, we have
\[
\mbox{Cov}(y_i, \sum_{\ell\in \alpha_i} \omega_{i\ell} x_{\ell}) 
= \begin{cases} \omega_{i\ell} \mbox{Cov}(y_i, x_i) & \text{for } \ell =\iota_i \text{ given } \iota_i \in \alpha_i \\ 0 & \text{if } \iota_i \not \in \alpha_i \end{cases}
\]
Thus, even though one does not know which links are the matches, as long as the links can cover a certain amount of matches, GREG estimation that makes appropriate use of the auxiliary data via the links still has the capacity to improve the HT-estimator. 

It is more difficult to draw general conclusions regarding the relative efficiency of the different types of GREG estimator. Take for instance the PI-GREG and the SRI-GREG estimators. The population GREG residual is $e_{iz} = y_i - z_i^{\top} B_z$ under the former, where $z_i = \sum_{\ell \in \alpha_i} \omega_{i\ell} x_{\ell}$, the residual is $e_{i\omega} = y_i - x_{i\omega}^{\top} B_{\omega}$ under the latter, where $x_{i\omega} = \sum_{\ell \in \alpha_i} \omega_{i\ell} x_{\ell}$. Although both $z_i$ and $x_{i\omega}$ are weighted sums of $x_{\ell}$ over the same $\alpha_i$, the weights sum to 1 for any $\ell$ in $A$ for the former whilst they sum to 1 for $i$ in $U$ for the latter. The relative magnitude of the residuals cannot be determined generally for each $i\in U$ \emph{on its own}, because it also depends on how the other units are linked. In the next section, we shall use a simulation study to explore the relative efficiency of the different estimators.

\section{Simulation study} \label{simulation}

\subsection{Set-up}

First, we generate a set of values $\{ (y_i, x_i) : i\in U\}$, where 
\[
y_ i =1+5 x_i + \epsilon_i \quad\text{and}\quad x_i \sim \mbox{Uniform}(0,1)
\quad\text{and}\quad \epsilon_i \sim \mbox{Normal}(0,\sigma_i^2)
\]
The model variance of $\epsilon_i$ is $\sigma_i = \sigma x_i^{\gamma}$ for $0\leq \gamma \leq 1$. We present the results under simple random sampling without replacement, where the sample size is $n$. Size-related unequal probability sampling does not yield any extra insight regarding the relative efficiency of the different estimators, because their relative merits are chiefly determined by how the population links $L$ are distributed over $U\times A$, regardless the sampling design.

We let $A = U$, so that we can easily calculate the \emph{ideal} GREG estimator \eqref{GREG}. The population matches and links are generated according to the parameters below. 
\begin{itemize}[leftmargin=5mm]
\item Let $p_d$ be the proportion of population units with $d_i =d$, where $d=1, 2, 3$ and $\sum_d p_d = 1$. For example, if $p = (0.4, 0.3, 0.3)$, then 40\% of the units in $U$ have only one link to $A$, 30\% of them have two links and the rest 30\% have 3 links.

\item Let $p_M$ be the proportion of units in $U$ that have a match in $A$, where $p_1 < p_M < 1$. By setting $p_M < 1$, one can emulate the general situation where $U$ and $A$ are not one-one correspondent in terms of the matches, and the ideal GREG estimator that uses all $\{ (y_i, x_i) : i\in s\}$ is unattainable in reality even if one knew all the matches. We let all the unique links be matches, the other $N(p_M - p_1)$ units with matches are randomly selected, independently of whether a unit has 2 or 3 links. For each $i\in U$ with $d_i >1$, all its false links are randomly selected from $A\setminus \{ i\}$.  

\item Let $p_{ML}$ be the proportion of units in $N$ whose matches are identified as the best links, where $p_1 \leq p_{ML} \leq p_M$. Setting $p_{ML} =p_M$ implies that the best-link choice is perfect given $(M, L)$, in which case the SBL-GREG estimator \eqref{SBL} reaches its maximum potential. Setting $p_{ML} =p_1$ means that all the known correct links are presented as the unique links. Using the SBL-GREG estimator is then unlikely to be a good option, because one could have obtained additional correct links among the $N(p_M - p_1)$ units just by guessing randomly. Thus, the SBL-GREG estimator improves as $p_{ML}$ varies from $p_1$ to $p_M$. The $N(p_{ML} - p_1)$ units with $d_i >1$ and correct best links are randomly chosen from the relevant $N(p_M - p_1)$ units. The best links for the rest $N(1-p_{ML})$ units are randomly chosen among the relevant false links in the respective $\alpha_i$.
\end{itemize}
Given each sample $s$, we calculate the following estimates and their variance estimates. 
\begin{itemize}[leftmargin=5mm]
\item The HT-estimator, and the ideal GREG estimator \eqref{GREG}, or simply Ideal.

\item The \emph{subsample} GREG-estimator, or simply Sub, which is only based on the sample units with $d_i =1$, i.e. with known correct links. This is a practical option, because in most applications of record linkage one can identify a subset of unique links that are virtually error-free, no matter how large or small this subset is in a given situation. 

\item The PI-GREG-estimator \eqref{PI} with multiplicity weights $\omega_{i\ell} = 1/m_{\ell}$ or unequal incidence weights as explained below, designated as PI-$m$ and PI-$q$, respectively. 

\item The SBL-GREG estimator \eqref{SBL}, or simply SBL, and the SRI-GREG estimator \eqref{SRI} with reverse incidence weights as explained below and designated as SRI-$q$.

\item The SLS-GREG estimator \eqref{SLS}, or simply SLS, with the same weights as SRI-$q$.
\end{itemize}

For SRI-$q$, the reverse incidence weight \eqref{ws} assigned to the best link is $\omega_{i\ell_i} =q$ in cases of $d_i>1$, where $0<q<1$, and $\omega_{i\ell} = (1-q)/(d_i -1)$ for the other links in $i\times \alpha_i$.
\begin{itemize}[leftmargin=5mm]
\item For a unit with $d_i =2$, setting $q=0.5$ would mean that one has no plausible guess which of the two links is more likely to be correct; for a unit with $d_i =3$, the indifferent choice would be $q=1/3$. For an easy presentation without unnecessary finesse, we simply set $q=0.4$ whether of $d_i =2$ or 3, which refers to a choice where the weights are more or less indifferent over the multiple links of unit $i$ in $U$. 

\item Of course, in cases where $p_{ML}$ is much higher than $p_1$, it is no longer reasonable to set $q=0.4$. To take advantage of the knowledge of linkage quality, one can raise the value of $q$, according to the proportion of units with correct best link given $d_i >1$, which is given by $(p_{ML}- p_1)/(1 - p_1)$. For example, if $(p_1, p_{ML}) = (0.2, 0.8)$, then setting $q$ around $(0.8 -0.2)/(1 -0.2) = 0.75$ is not an unnatural choice in practice. 
\end{itemize}

For the incidence weights \eqref{wp}, the multiplicity weight $1/m_{\ell}$ is the indifferent choice. For unequal weights of PI-$q$ when $m_{\ell} >1$, we proceed as follows: if the matched population unit is in $\beta_{\ell}$, assign the value $q$ to the match, where $0<q<1$, and $(1-q)/(m_{\ell} -1)$ to the other links in $\beta_{\ell}\times \ell$; otherwise, assign $q$ to a randomly selected link, and $(1-q)/(m_{\ell} -1)$ to the others. The value of $q$ can be large, if one has good knowledge of the linkage quality, such as when $p_{ML} = p_1$. Setting a lower value of $q$, e.g. $q=0.4$, emulates a situation where one has only vague ideas about the linkage quality. Notice that, given how the population links $L$ are generated above, the range of $m_{\ell}$ over $A$ is greater than than that of $d_i$ over $U$, although a large majority of records in $A$ still have $m_{\ell}$ between 0 and 3.  

Finally, based on $K$ independent samples, the Monte Carlo expectation and variance of an estimator, generically denoted by $t_{(k)}$ for $k=1, ..., K$, are given by
\[
\bar{t} = \frac{1}{K} \sum_{k=1}^K t_{(k)} \qquad\text{and}\qquad v(t) = \frac{1}{K-1} \sum_{k=1}^K (t_{(k)} - \bar{t})^2
\]
We obtain the MSE of the estimator accordingly. Moreover, the Monte Carlo expectation of the associated variance estimator, denoted by $\nu_{(k)}$ for $k=1, ..., K$, is given as 
\[
\bar{\nu}(t) = \frac{1}{K} \sum_{k=1}^K \nu_{(k)} 
\]

\subsection{Results} \label{results}

The population values of $y_i$ are generated with $\sigma_i \equiv 1.5$, where $N=5000$. The sample size is $n=100$. Let the population mean $\bar{Y}$ be the target of estimation. 

For the results in all the tables, SE is the square root of $v(t)$ of the corresponding estimator and ESE the square root of $\bar{\nu}(t)$. The variance estimator of an estimator works well, if its SE and ESE are close to each other. The relative efficiency (RE) of an estimator is given by the ratio between its variance and that of the HT-estimator, whereas RMSE designates the ratio of their MSEs. The bias of an estimator is small compared to its variance if its RMSE and RE are close to each other. 

The columns in all the tables refer to the different estimators by their shorthands given above. We simply set $c_i \equiv 1$ for GREG over $s$, and $c_{i\ell} \equiv 1$ for GREG over $L_s$.

\subsubsection{Low linkage quality}

Table \ref{tab-low1} provides a set of results in a situation where the linkage quality is very low. We have $p = (0.2, 0.4, 0.4)$, such that one is only confident about 20\% of the units whose links are matches. Next, we have $p_M =0.4$, such that the matches are missed from the relevant links for 60\% of the population units. The choice of the best link deteriorates as $p_{ML}$ decreases from $p_M$ to $p_1$. We set $q=0.4$ for all the results in Table \ref{tab-low1}, which is not unreasonable given the low linkage quality here. 

\begin{table}[ht]
\begin{center}
\caption{Results given low linkage quality, $N=5000$, $n=100$, $K=5000$}
\begin{tabular}{l ccc cc cc c}\hline\hline
& \multicolumn{8}{c}{$p = (0.2, 0.4, 0.4)$, $p_M = 0.4$, $p_{ML} = 0.4$, $q =0.4$} \\ 
& HT & Ideal & Sub & PI-$m$ & PI-$q$ & SBL & SRI-$q$ & SLS \\ \hline 
SE & 0.204 & 0.148 & 0.328 & 0.204 & 0.204 & 0.197 & 0.198 & 0.201\\ 
ESE & 0.203 & 0.146 & 0.327 & 0.201 & 0.201 & 0.194 & 0.195 & 0.199\\ 
RE & 1 & 0.525 & 2.595 & 1.002 & 1.001 & 0.930 & 0.939 & 0.968\\ 
RMSE & 1 & 0.525 & 2.596 & 1.002 & 1.001 & 0.931 & 0.940 & 0.968\\  \hline
& \multicolumn{8}{c}{$p = (0.2, 0.4, 0.4)$, $p_M = 0.4$, $p_{ML} = 0.3$, $q =0.4$} \\ 
& HT & Ideal & Sub & PI-$m$ & PI-$q$ & SBL & SRI-$q$ & SLS \\ \hline 
SE & 0.205 & 0.148 & 0.332 & 0.205 & 0.205 & 0.203 & 0.199 & 0.203\\ 
ESE & 0.204 & 0.146 & 0.325 & 0.202 & 0.202 & 0.198 & 0.195 & 0.200\\ 
RE & 1 & 0.522 & 2.623 & 0.999 & 1.001 & 0.977 & 0.943 & 0.977\\ 
RMSE & 1 & 0.522 & 2.625 & 0.999 & 1.001 & 0.979 & 0.944 & 0.978\\  \hline
& \multicolumn{8}{c}{$p = (0.2, 0.4, 0.4)$, $p_M = 0.4$, $p_{ML} = 0.2$, $q =0.4$} \\ 
& HT & Ideal & Sub & PI-$m$ & PI-$q$ & SBL & SRI-$q$ & SLS \\ \hline 
SE & 0.202 & 0.146 & 0.332 & 0.202 & 0.202 & 0.201 & 0.194 & 0.198\\ 
ESE & 0.203 & 0.147 & 0.325 & 0.201 & 0.201 & 0.200 & 0.194 & 0.199\\ 
RE & 1 & 0.522 & 2.691 & 1.000 & 0.998 & 0.984 & 0.924 & 0.959\\ 
RMSE & 1 & 0.522 & 2.692 & 1.000 & 0.998 & 0.985 & 0.924 & 0.960\\ \hline
\end{tabular} \label{tab-low1}
\end{center}
\end{table} 

First, since HT and Ideal do not depend on $L$, their Monte Carlo variance and MSE all have the same expectations in Table \ref{tab-low1}, such that the variations across the three blocks reflect directly the magnitudes of the Monte Carlo simulation errors. It is seen that the results are reliable within a range of $10^{-2}$. Although the variation is greater for Sub, as it is only based on about 20 sample units, it is clearly the least efficient estimator here.

Next, as can be expected, the performance of SBL worsens as $p_{ML}$ decreases. Its RE is about 1 when $p_{ML} = p_1 = 0.2$. However, since $p_{ML}$ is unlikely to be as low as $p_1$ in practice, one can still expect it to be slightly more efficient than HT.

Given constant $q=0.4$ in Table \ref{tab-low1}, only small variations of the results can be detected across the three blocks, regarding the variance and MSE of each of the other estimators. It follows that the population variations of $(M,L)$ and best links across the blocks do not affect the following conclusions based on these results. Using the incidence weights, PI-$m$ and PI-$q$ do not yield any gains over HT, although they are more difficult and costly to implement because they require the knowledge of $L$. Between SRI-$q$ and SLS, both based on the reverse incidence weights, the former is somewhat more efficient. In particular, SRI-$q$ is able to improve HT, even when $p_{ML}$ is as low as $p_1$, whereas it is about as efficient as SBL when $p_{ML} = 0.4$ and the latter is at its best. 

Comparing RE and RMSE, one can see that the bias is negligible compared to the variance for SBL, SRI-$q$ and SLS that only require the sample links $L_s$. Finally, comparing SE and ESE, one can see that the variance estimators work well in all the cases.

\begin{table}[ht]
\begin{center}
\caption{Results given low linkage quality, $N=5000$, $n=100$, $K=5000$}
\begin{tabular}{l ccc cc cc c}\hline\hline
& \multicolumn{8}{c}{$p = (0.2, 0.4, 0.4)$, $p_M = 0.8$, $p_{ML} = 0.8$, $q =0.4$} \\ 
& HT & Ideal & Sub & PI-$m$ & PI-$q$ & SBL & SRI-$q$ & SLS \\ \hline 
SE & 0.206 & 0.149 & 0.333 & 0.199 & 0.198 & 0.171 & 0.186 & 0.192\\ 
ESE & 0.204 & 0.146 & 0.324 & 0.194 & 0.194 & 0.168 & 0.183 & 0.190\\ 
RE & 1 & 0.524 & 2.622 & 0.933 & 0.932 & 0.691 & 0.818 & 0.876\\ 
RMSE & 1 & 0.524 & 2.624 & 0.933 & 0.932 & 0.691 & 0.822 & 0.878\\  \hline
& \multicolumn{8}{c}{$p = (0.2, 0.4, 0.4)$, $p_M = 0.8$, $p_{ML} = 0.8$, $q =0.7$} \\ 
& HT & Ideal & Sub & PI-$m$ & PI-$q$ & SBL & SRI-$q$ & SLS \\ \hline 
SE & 0.206 & 0.149 & 0.325 & 0.198 & 0.193 & 0.172 & 0.174 & 0.191\\ 
ESE & 0.204 & 0.146 & 0.326 & 0.195 & 0.189 & 0.169 & 0.172 & 0.190\\ 
RE & 1 & 0.519 & 2.481 & 0.924 & 0.872 & 0.694 & 0.716 & 0.861\\ 
RMSE & 1 & 0.519 & 2.482 & 0.924 & 0.873 & 0.697 & 0.719 & 0.861\\ \hline
& \multicolumn{8}{c}{$p = (0.2, 0.4, 0.4)$, $p_M = 0.8$, $p_{ML} = 0.2$, $q =0.4$} \\ 
& HT & Ideal & Sub & PI-$m$ & PI-$q$ & SBL & SRI-$q$ & SLS \\ \hline 
SE & 0.206 & 0.149 & 0.333 & 0.199 & 0.199 & 0.205 & 0.186 & 0.194\\ 
ESE & 0.204 & 0.147 & 0.328 & 0.195 & 0.195 & 0.201 & 0.182 & 0.191\\ 
RE & 1 & 0.519 & 2.599 & 0.931 & 0.930 & 0.983 & 0.815 & 0.883\\ 
RMSE & 1 & 0.519 & 2.604 & 0.931 & 0.930 & 0.983 & 0.818 & 0.884\\ \hline \hline 
\end{tabular} \label{tab-low2}
\end{center}
\end{table} 

Table \ref{tab-low2} provides another set of results, where $p$ remains the same but $p_M$ is increased to 0.8, such that 80\% of the population units now have their matches included in the links, although one can only be confident that about 20\% of the sample units are linked correctly. This provides a scenario where one can possibly have good knowledge of the linkage quality, although the available linkage key variables are rather noisy. Sub cannot improve given the same $p_1$. SBL is much better when $p_{ML} = p_M$, where it uses correctly matched auxiliary data for 80\% of the units, and its RE is about 0.69 in Table \ref{tab-low2} compared to 0.93 in Table \ref{tab-low1} when $p_{ML} = p_M$. But the gain easily evaporates as $p_{ML}$ decreases towards $p_1$. 
Although the results for PI-$m$ and PI-$q$ are better than before, they are still dominated by SRI-$q$ and SLS based on the reverse incidence weights, and the same pattern as before remains of the relative merits of the latter two. Again, the bias is negligible compared to the variance and the variance estimators work well in all the cases.

A notice is worthwhile regarding the second block of results in Table \ref{tab-low2}. Now that $p_{ML} = 0.8$ is much higher than $p_1 =0.2$, it is no longer reasonable to set $q=0.4$, where the weights are more or less indifferent over the multiple links. To take advantage of the good knowledge of linkage quality, one can raise the value of $q$. Setting $q=0.7$ is not hard to justify here, given $(p_{ML}- p_1)/(1 - p_1) = 0.75$. While this clearly improves the results for SRI-$q$, where the RE is 0.71 against 0.82 given $q=0.4$ in the first block, it does not have any noteworthy effect for SLS. In the case of SLS, GREG is over $L_s$ instead of $s$, and it seems more difficult to assign the unequal weights sensibly for this estimator. PI-$q$ is also clearly better given larger $q$, where its RE is 0.87 compared to 0.93 in the first block where $q=0.4$, although the improvement is not as large as for SRI-$q$. 

Finally, GREG estimation is much more efficient than the HT-estimator, at least in these results, even when one can only be certain that about 20\% of the sample units are correctly matched, as long as $L$ covers a large part of $M$. For instance, the SRI-GREG estimator achieves about 20\% variance reduction in the last block, only based on more or less indifferent reverse incidence weights for the units with multiple links.

\subsubsection{Better linkage quality}

Two more sets of results are given in Table \ref{tab-high}, given better linkage quality than above. For the first two blocks of results, we have $p = (0.4, 0.3, 0.3)$ and $p_M = 0.9$, such that 90\% of the population units have matches among the links, although one is only certain about nearly half of them. This will be referred to as the \emph{medium} linkage quality scenario. For the last two blocks, we have $p = (0.8, 0.1, 0.1)$ and $p_M = 0.98$, such that only 2\% of the population units have missing matches in $L$, and nearly 80\% of the matches are given as unique links. This will be referred to as the \emph{high} linkage quality scenario. 

The following features are essentially the same as the results in Tables \ref{tab-low1} and \ref{tab-low2} given low linkage quality. In all the cases, the bias is negligible compared to the variance and the variance estimators work well. PI-$m$ and PI-$q$ using incidence weights are still largely dominated by SRI-$q$ and SLS using reverse incidence weights. For the former two estimators, PI-$q$ can improve PI-$m$ given good knowledge of the linkage quality, i.e. when $p_{ML} = p_M$; for the latter two estimators, SRI-$q$ is still better than SLS.  
   
\begin{table}[ht]
\begin{center}
\caption{Results given medium-high linkage quality, $N=5000$, $n=100$, $K=5000$}
\begin{tabular}{l ccc cc cc c}\hline\hline
& \multicolumn{8}{c}{$p = (0.4, 0.3, 0.3)$, $p_M = 0.9$, $p_{ML} = 0.9$, $q =0.7$} \\ 
& HT & Ideal & Sub & PI-$m$ & PI-$q$ & SBL & SRI-$q$ & SLS \\ \hline 
SE & 0.205 & 0.149 & 0.232 & 0.194 & 0.183 & 0.160 & 0.164 & 0.184\\ 
ESE & 0.203 & 0.146 & 0.233 & 0.191 & 0.181 & 0.157 & 0.160 & 0.181\\ 
RE & 1 & 0.532 & 1.278 & 0.900 & 0.800 & 0.612 & 0.638 & 0.805\\ 
RMSE & 1 & 0.532 & 1.279 & 0.901 & 0.800 & 0.615 & 0.641 & 0.806\\  \hline
& \multicolumn{8}{c}{$p = (0.4, 0.3, 0.3)$, $p_M = 0.9$, $p_{ML} = 0.65$, $q =0.4$} \\ 
& HT & Ideal & Sub & PI-$m$ & PI-$q$ & SBL & SRI-$q$ & SLS \\ \hline 
SE & 0.204 & 0.147 & 0.229 & 0.191 & 0.192 & 0.183 & 0.173 & 0.184\\ 
ESE & 0.203 & 0.146 & 0.233 & 0.190 & 0.191 & 0.181 & 0.173 & 0.183\\ 
RE & 1 & 0.519 & 1.262 & 0.878 & 0.891 & 0.804 & 0.723 & 0.810\\ 
RMSE & 1 & 0.519 & 1.263 & 0.879 & 0.892 & 0.804 & 0.723 & 0.811\\ \hline \hline
& \multicolumn{8}{c}{$p = (0.8, 0.1, 0.1)$, $p_M = 0.98$, $p_{ML} = 0.98$, $q =0.9$} \\ 
& HT & Ideal & Sub & PI-$m$ & PI-$q$ & SBL & SRI-$q$ & SLS \\ \hline 
SE & 0.201 & 0.145 & 0.164 & 0.174 & 0.155 & 0.148 & 0.149 & 0.162\\ 
ESE & 0.204 & 0.146 & 0.165 & 0.175 & 0.155 & 0.149 & 0.150 & 0.164\\ 
RE & 1 & 0.526 & 0.667 & 0.753 & 0.593 & 0.547 & 0.548 & 0.651\\ 
RMSE & 1 & 0.526 & 0.675 & 0.753 & 0.593 & 0.549 & 0.549 & 0.651\\ \hline
 & \multicolumn{8}{c}{$p = (0.8, 0.1, 0.1)$, $p_M = 0.98$, $p_{ML} = 0.89$, $q =0.4$} \\ 
& HT & Ideal & Sub & PI-$m$ & PI-$q$ & SBL & SRI-$q$ & SLS \\ \hline
SE & 0.205 & 0.149 & 0.166 & 0.179 & 0.182 & 0.162 & 0.158 & 0.165\\ 
ESE & 0.204 & 0.146 & 0.165 & 0.174 & 0.178 & 0.160 & 0.156 & 0.164\\ 
RE & 1 & 0.532 & 0.660 & 0.763 & 0.793 & 0.625 & 0.596 & 0.652\\ 
RMSE & 1 & 0.532 & 0.666 & 0.764 & 0.794 & 0.626 & 0.596 & 0.654\\  \hline \hline
\end{tabular} \label{tab-high}
\end{center}
\end{table} 

Some additional points are worth noting. Sub becomes more efficient than HT given high linkage quality where $p_1 = 0.8$. However, just like Ideal, it is infeasible in reality, because one cannot be sure if the auxiliary total $X_A$ is equal to $\sum_{i\in U} x_i$. We have set $A=U$ here to ensure the two are equal, only so that these two estimators can be easily calculated, in order to provide references for the performance of the GREG estimators developed in this paper. In the high linkage quality scenario, the RE is about 0.66 for Sub, so that it is dominated by SBL and SRI-$q$, because they use the additional auxiliary information for the population units with multiple links. Moreover, the RE of SBL and SRI-$q$ are 0.55 when $p_{ML} = p_M = 0.98$ and $q=0.9$, which is about the same as 0.53 for Ideal, suggesting that the auxiliary information is almost fully utilised. 

The settings $(p_M, p_{ML}) = (0.4, 0.65)$ and $(0.8, 0.89)$ emulate random selection of the best link among the available links, where $p_{ML} = p_1 + p_2/2 + p_3/3$ in the first case and approximately so in the second case. For either case, the reverse incidence weights given $q=0.4$ are also more or less indifferent over the available links. The RE is 0.72 for SRI-$q$ against 0.80 for SBL in the medium quality scenario, and it is 0.60 for SRI-$q$ against 0.63 for SBL in the high quality scenario. Thus, it is possible to improve the simplistic SBL-GREG estimator through the choice of reverse incidence weights, even though one has no knowledge about the correct link given multiple links.

As discussed before, when $p_{ML} = p_M$ and the choice of best link is perfect for the given $(M,L)$, it is reasonable to use a higher value of $q$. The choice of $q=0.7$ is not unnatural in the medium quality scenario where $(p_{ML} - p_1)/(1 -p_1) = 0.83$, and similarly for $q=0.9$ in the high quality scenario where $(p_{ML} - p_1)/(1 -p_1) = 0.9$. The RE is 0.64 for SRI-$q$ against 0.61 for SBL in the medium quality scenario, and it is 0.55 for SRI-$q$ against 0.55 for SBL in the high quality scenario.
This suggests that it is not difficult for SRI-$q$ to make most out of the good knowledge of linkage quality, characterised as $p_{ML} = p_M$, by assigning appropriate reverse incidence weight to the best link accordingly. 

Meanwhile, changing the value of $q$ has basically no affect on SLS in either scenario, indicating again that it may be more difficult to assign unequal weights sensibly for this estimator. Raising the value of $q$ does much good for PI-$q$, where its RE is improved from 0.89 to 0.80 in the medium quality scenario, and from 0.79 to 0.59 in the high quality scenario. Nevertheless, it is dominated by SRI-$q$ and SBL. The reason is that $m_{\ell}$ varies more over $A$ than $d_i$ over $U$, due to directional linkage from $U$ to $A$, where one can easily control the latter range but not the former. Thus, when it is possible to link the entire $U$ and $A$, one may be able to improve PI-$q$ by adopting a two-way linkage method.

\section{Conclusions and final remarks}\label{final}

Three types of GREG estimators are developed given imperfectly matched auxiliary data, where the standard GREG estimator is inapplicable. Simulation results show that they can improve the estimation efficiency, compared to the HT-estimator that ignores the auxiliary information, even when the linkage quality is as low as that given for Table \ref{tab-low1}. Other simulations, omitted here to save space, yield results that are consistent with those reported in Section \ref{results}, as the sample size varies between $n=30$ and $n=1000$, or when the regression model variance is as heterogenous as $\sigma_i = 2 x_i$.

The first type of PI-GREG estimator \eqref{PI} is design-consistent, but costly or impossible to implement, because it is based on the incidence weights \eqref{wp} that require one to link the entire population and auxiliary database. Moreover, it lacks efficiency compared to the other two types of GREG estimators in the simulation study. However, as discussed at the end of Section \ref{results}, one may be able to improve it by adopting a two-way linkage method, provided one can obtain good knowledge of the linkage quality.  

The other two types of GREG estimators are more practical, as they are based on the reverse incidence weights \eqref{ws}, for which one only needs to link the sample to the auxiliary database. A special case is the simplistic best-link estimator \eqref{SBL} that may be relevant for secondary users who have no access to the auxiliary database or the record linkage procedure. The additional assumptions for these estimators to be design-consistent can be tested given the observed sample and links, and MSE can be used as the uncertainty measure instead of sampling variance if the bias is not negligible. The simulation study demonstrates that the SRI-estimator \eqref{SRI} can more easily be made efficient, compared to the SLS estimator \eqref{SLS}, through the weights assigned to the best links. 

In summary, the SRI-GREG estimator is easy to implement and, at this stage, is often the most efficient given sensible choices of the reverse incidence weights. Future research will hopefully provide better theoretical insights to the relative efficiency of the estimators, and it will be intriguing to see whether the PI-GREG and SLS-GREG estimators can be made more competitive, or if there are potentially other effective approaches to devising GREG estimators given imperfectly matched auxiliary data.


\begin{thebibliography}{99}

\bibitem{Breidt2018ModelAssistedSE} Breidt, J., Opsomer, J.D. and Huang, C.-M. (2018). Model-Assisted Survey Estimation with Imperfectly Matched Auxiliary Data, in \textit{Predictive Econometrics and Big Data}, Springer International Publishing. 

\bibitem{Chambers:09} Chambers R. (2009). Regression analysis of probability-linked data. \textit{Official Statistics Research Series, Vol. 4.} Statistics New Zealand. 

\bibitem{chambers2020} Chambers, R.L. and A.D. da Silva (2020). Improved secondary analysis of linked data: a framework and an illustration. \textit{Journal of the Royal Statistical Society: Series A}. {\bf 183}, 37--59. doi: \url{10.1111/rssa.12477}

\bibitem{christen2012} Christen, P. (2012). A survey of indexing techniques for scalable record linkage and deduplication. \textit{ISEE Transactions on Knowledge and Data Engineering}, {\bf 24}. 

\bibitem{harron2015} Harron, K., Goldstein, H. and Dibben, C. (2015). \textit{Methodological Developments in Data Linkage}. Wiley.

\bibitem{herzog2007} Herzog, T.N., Scheuren, F.J. and Winkler, W.E. (2007). \textit{Data Quality and Record Linkage Techniques}. Springer.

\bibitem{fellegi1969} Fellegi,I .P. and Sunter, A.B. (1969). A theory for record linkage, \emph{Journal of the American Statistical Society},  {\bf 64}, 1183--1210.  

\bibitem{ll2005} Lahiri, P. and Larsen, M.D. (2005). Regression analysis with linked data. \textit{Journal of the American Statistical Association}, {\bf 100}, 222--230.

\bibitem{sarndal1992} S\"{a}rndal, C.-E., Swensson, B. and Wretman, J. (1992). \textit{Model Assisted Survey Sampling.} New York: Springer-Verlag.

\bibitem{lcz2019} Zhang, L.-C. and Tuoto, T. (2019). Linkage-data linear regression. \textit{Journal of the Royal Statistical Society: Series A}. \textit{Revised.}

 \end{thebibliography}
\end{document}